\pdfoutput=1
\documentclass[11pt,aps,pre,superscriptaddress,byrevtex,bibnotes,footinbib,showpacs,a4paper,longbibliography,notitlepage]{revtex4-1}
\usepackage{microtype}
\usepackage{graphicx}
\usepackage{amsmath}
\usepackage{amssymb}
\usepackage{color}
\definecolor{myblue}{rgb}{0.153,0.322,0.706}
\usepackage[colorlinks,linkcolor=myblue,urlcolor=myblue,citecolor=myblue,breaklinks=true]{hyperref}

\DeclareMathOperator{\Tr}{Tr}

\renewcommand{\emph}{\textit}
\newcommand{\be}{\begin{equation}}
\newcommand{\ee}{\end{equation}}
\newcommand{\ra}{\rightarrow}
\newcommand{\cL}{\mathcal{L}}
\newcommand{\cK}{\mathcal{K}}
\newcommand{\cB}{\mathcal{B}}
\newcommand{\reals}{\mathbb{R}}
\newcommand{\p}{\partial}
\newcommand{\transp}{\mathsf{T}} 
\newcommand{\id}{\mathbb{I}}
\newcommand{\bs}{\boldsymbol}
\newcommand{\bx}{\bs{x}}
\newcommand{\by}{\bs{y}}
\newcommand{\bv}{\bs{v}}
\newcommand{\bX}{\bs{X}}
\newcommand{\bW}{\bs{W}}
\newcommand{\bF}{\bs{F}}
\newcommand{\bJ}{\bs{J}}
\newcommand{\bsn}{\bs{\nabla}}
\newcommand{\ta}{\bar a}
\newcommand{\tbF}{\bs{\widetilde F}}
\newcommand{\tp}{\tilde p}
\newcommand{\tbJ}{\bs{\widetilde J}}
\newcommand{\ce}{\bar e}
\newcommand{\E}{\mathbb{E}}
\newcommand{\covm}{C}
\newcommand{\obsQ}{Q}
\newcommand{\obsm}{\Gamma}
\newcommand{\obsv}{\bs{\eta}}
\newcommand{\ricmat}{B}
\newcommand{\ricval}{b}
\newcommand{\work}{\mathcal{W}_T}
\newcommand{\ep}{\mathcal{E}_T}
\newcommand{\dprod}[2]{\left\langle #1, #2\right\rangle}

\begin{document}
\title{Dynamical large deviations of linear diffusions}

\author{Johan du Buisson}
\email{johan.dubuisson@gmail.com}
\affiliation{Department of Physics, Stellenbosch University, Stellenbosch 7600, South Africa}

\author{Hugo Touchette}
\email{htouchette@sun.ac.za}
\affiliation{Department of Mathematical Sciences, Stellenbosch University, Stellenbosch 7600, South Africa}

\date{\today}

\begin{abstract}
Linear diffusions are used to model a large number of stochastic processes in physics, including small mechanical and electrical systems perturbed by thermal noise, as well as Brownian particles controlled by electrical and optical forces. Here, we use techniques from large deviation theory to study the statistics of time-integrated functionals of linear diffusions, considering three classes of functionals or \emph{observables} relevant for nonequilibrium systems which involve linear or quadratic integrals of the state in time. For these, we derive exact results for the scaled cumulant generating function and the rate function, characterizing the fluctuations of observables in the long-time limit, and study in an exact way the set of paths or effective process that underlies these fluctuations. The results give a complete description of how fluctuations arise in linear diffusions in terms of effective forces that remain linear in the state or, alternatively, in terms of fluctuating densities and currents that solve Riccati-type equations. We illustrate these results using two common nonequilibrium models, namely, transverse diffusions in two dimensions involving a non-conservative rotating force, and two interacting particles in contact with heat baths at different temperatures.
\end{abstract}

\maketitle

\section{Introduction}

Stochastic differential equations (SDEs) are widely used in science and engineering to model the dynamics of systems driven by both deterministic forces and external noise sources \cite{gardiner1985,risken1996,jacobs2010,pavliotis2014}. In many cases, the force acting on a system can be taken or approximated to be linear in the state, giving rise to linear SDEs, which are also referred to as linear diffusions, linear Langevin equations or Ornstein--Uhlenbeck processes. These are used in physics to model many different systems evolving close to fixed points or in weak force regimes, including small mechanical systems perturbed by thermal noise \cite{volpe2006,geitner2017,gomez2010,ciliberto2017}, and Brownian particles manipulated by electrical or laser fields \cite{ashkin1997,ritort2008,ciliberto2017}. Because they are exactly solvable, linear SDEs are also used as basic models of nonequilibrium systems to study the effect of non-conservative forces, temperature gradients, and the breaking of time-reversal symmetry, in general, on the steady state of these systems, determined by their stationary density and current \cite{zia2007,weiss2007,seifert2012}.

In this work, we study the statistics of \emph{dynamical observables} of linear SDEs, defined as time-integrated functionals of the paths or trajectories of an SDE. These quantities are related in physics to thermodynamic quantities, such as the work done on a system over time or the heat exchanged with a bath, and thus play a prominent role when investigating the efficiency of control and biological processes that fluctuate at the micro and meso scales \cite{sekimoto2010,seifert2012,peliti2021}. The study of these fluctuations using techniques from large deviation theory \cite{touchette2009,harris2013,touchette2017,jack2020} has led in recent years to many general results and insights about the physics of nonequilibrium systems, related to fluctuation symmetries \cite{gallavotti1995,kurchan1998,lebowitz1999,harris2007}, fluctuation phase transitions \cite{garrahan2007,garrahan2009,speck2011,espigares2013,bunin2013,tsobgni2016b,lazarescu2017}, and thermodynamic uncertainty relations or bounds connecting the variance of current fluctuations to dissipation \cite{barato2015b,pietzonka2015,gingrich2016,gingrich2017,li2019}. 

Another important insight coming from large deviation theory is that fluctuations of observables in nonequilibrium processes arise from density and current fluctuations that organise themselves in an ``optimal'' way so as to minimize a certain cost or loss \cite{derrida2007,bertini2002,bertini2015b,maes2008a,maes2008b,barato2015,hoppenau2016}, similarly to noise-driven transitions in chemical systems which are known to follow optimal ``pathways'' \cite{freidlin1984,wales2004,vanden-eijnden2006}. This observation has proven useful for understanding the transport properties of many nonequilibrium systems, including interacting particle systems driven in nonequilibrium states by boundary reservoirs \cite{derrida2007,bertini2002,bertini2015b}, and generalizes at the level of fluctuations the idea that the knowledge of the stationary density \emph{and} the current of a Markov process is sufficient to completely characterise its stationary state \cite{zia2007}. 

This description of nonequilibrium systems in terms of density and current fluctuations is appealing physically, but cannot be developed easily in practice because determining the large deviation functions or potentials that describe these fluctuations requires that we solve non-trivial spectral or optimization problems \cite{touchette2017}, whose dimension increases with the size of the system or model considered. For this reason, there are only a few models for which these functions can be calculated exactly, including lattice jump processes \cite{garrahan2007,garrahan2009,speck2011,lecomte2005,lecomte2007,jack2014}, low-dimensional SDEs \cite{mehl2008,angeletti2015,tsobgni2016,tsobgni2018}, as well as simple interacting particle systems, such as the one-dimensional exclusion process \cite{derrida1998,derrida2002,derrida2003,lazarescu2011,mallick2015} or the zero-range process \cite{harris2005,grosskinsky2008,villavicencio2012,hirschberg2015}, which can also be solved analytically in the macroscopic limit using field-theory techniques from the macroscopic fluctuation theory \cite{bertini2015b}.

Here, we derive analytical results for the dynamical large deviations of linear SDEs, showing that this type of process admits exact solutions for three classes of observables, defined, respectively, as linear integrals of the state in time, quadratic integrals of the state, and stochastic integrals involving a linear function of the state multiplied by the increment of the process in time. Each of these arises naturally when dealing with linear SDEs, as shown in the next sections. The third class of observables, in particular, arises when defining thermodynamic quantities, such as the nonequilibrium work or the entropy production, related to the current.

Large deviations of linear SDEs have been studied for specific examples of linear SDEs and observables, notably, quadratic observables \cite{benitz1990,bercu1997,bryc1997,gamboa1999,landais1999,bercu2000,bercu2002,monthus2022}, the entropy production \cite{chernyak2006,turitsyn2007,jaksic2016,jaksic2017,monthus2022,costa2022}, the nonequilibrium work \cite{visco2006a,kwon2011,noh2013,noh2014}, energy currents in spin dynamics \cite{lecomte2005b}, and the heat exchanged by harmonic oscillators with a heat bath \cite{kundu2011b,sabhapandit2011,sabhapandit2012,pal2013,ciliberto2013,ciliberto2013b}. Our results unify and generalize these studies by considering linear SDEs in any dimension and by extending the class of observables considered to the three classes mentioned above. For these, we give explicit expressions, involving Riccati-type equations, for the scaled cumulant generating function and the rate function, which characterize the probability distribution of dynamical observables in the long-time limit. These two functions are important in physics, since they also determine symmetries in the distribution of observables, referred to as fluctuation relations \cite{gallavotti1995,kurchan1998,lebowitz1999,harris2007}, and sharp transitions between different fluctuation regimes \cite{garrahan2007,garrahan2009,speck2011,espigares2013,bunin2013,tsobgni2016b,lazarescu2017}.

Compared to previous studies, we also provide a complete description of the way fluctuations arise in terms of optimal density and current fluctuations, which modify the stationary density and current of the diffusion considered or, equivalently, in terms of an effective diffusion process that modifies the force or drift of the original diffusion \cite{chetrite2013,chetrite2014,chetrite2015,jack2015}. This effective process has been studied extensively in recent years for many examples of jump processes \cite{simon2009,popkov2010,jack2010b,grandpre2018,das2019} and diffusions \cite{angeletti2015,tsobgni2016,tsobgni2018,buisson2020b,tizon2019,yan2022} used as models of nonequilibrium systems, and has been shown in this context to be useful for understanding transitions between different fluctuation regimes, among other phenomena. One important property of this process that we uncover is that, for the three types of observables considered, the effective process is also described by a linear SDE, which means that it is characterized by a modified Gaussian stationary density and current corresponding, in the original SDE, to the optimal density and current fluctuations that give rise to an observable fluctuation. In this sense, these observables can be considered as ``closed'' or ``sufficient'' for the effective process to remain in the same SDE class as the original model.

The results that we obtain provide in the end a complete and exactly solvable framework for studying the large deviations of linear diffusions, which can be used to predict their steady-state and fluctuation properties, to determine from observed trajectories whether a system is reversible or irreversible, and, from a more applied perspective, to understand the convergence of Monte Carlo simulations \cite{coghi2021}. We believe that they can also be applied, beyond linear systems, to approximate the large deviation functions of nonlinear SDEs or nonlinear observables near fixed points of the corresponding noiseless dynamics, at least in the Gaussian regime of fluctuations, characterized by a mean and variance, which also follow from our results. This can be applied potentially to study the large deviations of more complex systems and to develop new numerical algorithms for computing large deviation functions. We comment on these issues in the concluding section.

To illustrate our results, we consider two linear models in two dimensions, commonly used in physics to describe nonequilibrium processes, namely, transverse diffusions driven by a non-conservative and thus nonequilibrium force that generates a rotating drift in the plane, and the so-called Brownian gyrator, which consists of two particles interacting via a linear (spring) force, put in contact with heat baths at different temperatures. For each of these models, we show for specific observables how the stationary density and current are changed when fluctuations are observed, and how these changes depend on non-conservative forces being applied or on external reservoirs. In some cases, the current can vanish at the fluctuation level, implying that an irreversible system can ``behave'' in a reversible way when it is observed to fluctuate in a specific way or direction.

\section{Model and large deviations}

\subsection{Linear SDEs}

The systems that we consider are underdamped diffusions or Langevin-type systems, described by the linear SDE
\be
d\bX (t) = -M\bX (t) dt + \sigma d\bW (t),
\label{SDE}
\ee
where $\bX (t)\in\reals^n$ is the system's state at time $t$, $M$ is the drift matrix defining the linear force or drift acting on $\bX (t)$, and $\bW (t)\in\reals^m$ is a vector of independent Brownian or Wiener motions acting as the noise source, which is multiplied by the noise matrix $\sigma$ of size $n\times m$. For the remaining, we assume that the diffusion matrix $D=\sigma\sigma^\transp$, where $\transp$ stands for the transpose, is invertible. 

Linear SDEs are used in physics to model many different systems, including nonequilibrium processes driven by temperature or chemical gradients \cite{gardiner1985}, small cantilever and torsion systems perturbed by thermal noise \cite{volpe2006,geitner2017,gomez2010,ciliberto2017}, Brownian particles manipulated by laser tweezers \cite{ashkin1997,ritort2008,ciliberto2017}, and electric circuits perturbed by Nyquist or artificial noise \cite{luchinsky1998,zon2004a,garnier2005}. In control theory, they are also widely used as exact or approximate models of feedback-controlled systems, forming the basis of the classical linear-quadratic-Gaussian control problem \cite{stengel1994,schulz2006,bechhoefer2021}. Naturally, part of the interest for linear SDEs comes from the fact that they can be solved exactly to express $\bX (t)$ as an integral of the noise. Moreover, the probability density $p(\bx,t)$ can be found exactly by solving the Fokker--Planck equation, yielding a Gaussian distribution at all times $t>0$ when the system is initialized at $\bX(0)=\bx_0$, characterized by a time-dependent mean and covariance matrix (see \cite[Sec.~3.7]{pavliotis2014}).

Here, we focus on the long-time behaviour of the SDE \eqref{SDE} by further assuming that the matrix $M$, which is not necessarily symmetric, is positive definite, meaning that its eigenvalues have positive real parts. In this case, it is known that the SDE is ergodic and, thus, has a unique stationary density, given explicitly \cite{pavliotis2014} by 
\be
p^*(\bx) = \sqrt{\frac{1}{(2 \pi)^n \det C}} \exp\left(-\frac{1}{2} \dprod{\bx}{\covm^{-1} \bx} \right),
\label{statden}
\ee
where $\covm$, the covariance matrix, satisfies the Lyapunov equation
\be
D = M\covm + \covm M^\transp.
\label{ricccov}
\ee
Here we use $\langle \bs{a}, \bs{b}\rangle$ to represent the standard vector inner product in $\reals^n$. In the long-time limit, the SDE is also characterized by a stationary current, defined in general by
\be
\bJ^*(\bx) = \bF(\bx) p^*(\bx) - \frac{1}{2} D\bsn p^*(\bx),
\label{current}
\ee
where $\bF(\bx)$ is the general force or drift entering in an SDE, which for $\bF(\bx)=-M\bx$ and $p^*(\bx)$ as given in \eqref{statden} yields the current
\be
\bJ^*(\bx) = H\bx p^*(\bx),
\label{statcurlin}
\ee
where
\be
H=\frac{D}{2} \covm^{-1} - M.
\label{currmat}
\ee

For the remaining, it is important to note that $p^*$ and $\bJ^*$ determine $\bF$ uniquely for a fixed $D$ via \eqref{current}, which means that their knowledge can be used to identify an SDE, whether linear or not. Moreover, the stationary current determines the reversibility of an SDE, that is, whether the probability of any given path over the time is the same as the probability of that path reversed in time \cite{zia2007}. If $\bJ^*(\bx)=\bs{0}$ for all $\bx$, then the SDE is reversible, describing in the long-time limit an equilibrium steady state, whereas if $\bJ^*(\bx)\neq 0$, then the SDE is irreversible and describes a nonequilibrium steady state violating the condition of detailed balance. 

For linear SDEs, we can distinguish two sources of nonequilibrium behavior: a non-symmetric drift matrix $M$, leading to a non-conservative $\bF$ that does not follow from the gradient of a potential, and a diffusion matrix $D$ not proportional to the identity matrix $\id$, related to heat baths with different temperatures or correlated noise sources. We study examples of these cases in Sec.~\ref{secapps}. Of course, a non-symmetric $M$ and $D\not\propto\id$ can still lead to an equilibrium state if they are such that $H=0$.

\subsection{Observables}

For a given linear SDE, we are interested in obtaining the long-time form of the probability density $p(A_T=a)$ of a dynamical observable $A_T$, which is a time-averaged function of the state $\bX (t)$. We consider, specifically, three classes of observables:
\begin{itemize}
\item Linear additive observables of the form
\be
A_T =\frac{1}{T}\int_0^T \dprod{\obsv}{\bX (t)} dt,
\label{linadd}
\ee
where $\obsv $ is an arbitrary vector in $\reals^n$;

\item Quadratic observables, defined as
\be
A_T = \frac{1}{T}\int_0^T \dprod{\bX (t)}{\obsQ\bX (t)} dt,
\label{quadadd}
\ee
where $\obsQ$ is assumed, without loss of generality, to be a symmetric $n\times n$ matrix;

\item Linear current-type observables, defined in terms of the increments of the SDE as
\be
A_T=\frac{1}{T}\int_0^T \obsm\bX (t)\circ d\bX (t),
\label{lincurobs}
\ee
where $\obsm$ is an arbitrary $n\times n$ matrix and $\circ$ denotes the scalar product taken according to the Stratonovich convention or calculus.
\end{itemize}

These observables are important in physics, as they include many quantities that can be measured in practice, such as the mechanical work done on a nonequilibrium process, the heat transferred in time between a system and its environment, and the entropy production, which is a measure of the irreversibility of stochastic processes \cite{sekimoto2010,seifert2012,peliti2021}. In control theory, the quadratic observable is also related to quadratic cost functions or Lagrangians that are minimized to determine the optimal control inputs in steady-state control systems \cite{stengel1994,schulz2006,bechhoefer2021}. We study specific examples in Sec.~\ref{secapps}, showing how the vector $\obsv $ and matrices $\obsQ$ and $\obsm$ are to be chosen depending on the observable considered. 

\subsection{Large deviation principle}

Finding the probability density of $A_T$ is difficult in general, even for linear SDEs. However, it is known from large deviation theory \cite{touchette2009,harris2013,touchette2017,oono1989,dembo1998,hollander2000} that this density often scales in the limit of large integration times $T$ according to 
\be
p(A_T = a)\approx e^{-T I(a)},
\ee
so the problem of finding $p(A_T=a)$ is simplified to the problem of finding the exponent $I(a)$, called the \emph{rate function}. The meaning of this approximation is that the dominant part of $p(A_T=a)$ as $T$ becomes large is a decaying exponential controlled by $I(a)$, so that corrections are sub-exponential in $T$. When this holds, $A_T$ is said to satisfy the \emph{large deviation principle} (LDP) with rate function $I(a)$ \cite{dembo1998}. Equivalently, $A_T$ is said to satisfy the LDP if the limit
\be
\lim_{T\ra\infty} -\frac{1}{T}\ln p(A_T=a) = I(a)
\ee
exists and yields a non-trivial rate function.

The rate function is generally convex for ergodic Markov processes, and has a unique minimum and zero, denoted here by $a^*$, which corresponds to the typical value of $A_T$ where $p(A_T=a)$ concentrates as $T\ra\infty$ \cite{dembo1998}. The LDP shows that this concentration is exponential with $T$, so that fluctuations of $A_T$ away from $a^*$ are exponentially unlikely with the integration time. 

The typical value $a^*$ also corresponds to the stationary expectation or mean of $A_T$, and so can be calculated from $p^*$ or $\bJ^*$, depending on the observable considered. For linear additive observables, we trivially have
\be
a^* = \int_{\reals^n}  \dprod{\obsv}{\bx}\, p^*(\bx) d\bx =0,
\label{meanadd}
\ee
since $p^*$ has zero mean, whereas for quadratic observables, $a^*$ is a modified second moment of $p^*$: 
\be
a^* = \int_{\reals^n} \dprod{\bx}{\obsQ\bx}\, p^*(\bx) d\bx = \Tr (\obsQ\covm).
\label{meanquad}
\ee
The result in both cases involves only $p^*$, and so $A_T$ is said to be a density-type observable. For the third class of observable considered, we have instead
\be
a^* = \int_{\reals^n}  \dprod{\obsm\bx}{ \bJ^*(\bx)} d\bx  = \Tr(\obsm^\transp HC),
\label{meancurr}
\ee
which explains why we refer to it as a current-type observable. In particular, $a^*=0$ for this observable if $\bJ^*=\bs{0}$.

\subsection{Large deviation functions}

To find the rate function of the three classes of observables defined above, we use the G\"artner--Ellis theorem \cite{dembo1998}, which expresses $I(a)$ in terms of another function $\lambda(k)$, known as the \emph{scaled cumulant generating function} (SCGF) and defined by
\be
\lambda(k) = \lim_{T\rightarrow \infty} \frac{1}{T} \ln \E[e^{kTA_T}],
\label{SCGF}
\ee
where $\E[\cdot ]$ denotes the expectation. Provided that $\lambda(k)$ exists and is differentiable, then $A_T$ satisfies the LDP and its rate function is given by the Legendre transform of the SCGF \cite{dembo1998}:
\be
I(a) = k_a a - \lambda(k_a),
\ee
with $k_a$ the unique root of
\be
\lambda'(k) = a. 
\label{duality}
\ee

The advantage of using the SCGF for obtaining the rate function is that the generating function of $A_T$ conditioned on $\bX (0)=\bx$, defined by 
\be
G_k(\bx,t) = \E[e^{ktA_t}|\bX (0)=\bx],
\ee
satisfies the linear equation
\be 
\p_t G_k(\bx,t) = \cL_k G_k(\bx,t),
\label{FK}
\ee
where $\cL_k$ is a modification of the generator of the SDE, known as the \emph{tilted generator}, which depends in our case on $M$, $D$ and the observable considered \cite{touchette2017}. We give in the next section the explicit expression of this operator as we consider the three observables individually. The linear equation defined by this operator is the well-known Feynman--Kac (FK) equation, which can be solved from the initial condition $G_k(\bx,0) = 1$ to obtain $G_k(\bx,t)$ and, in turn, $\lambda(k)$ by taking the long-time limit of this solution, which does not depend in general on the initial condition because of the ergodicity of the process.

Alternatively, we can use the fact that the FK equation is linear to expand $G_k(\bx,t)$ in a complete basis of bi-orthogonal eigenfunctions to obtain the SCGF, under mild conditions, from the dominant eigenvalue of $\cL_k$. The SCGF can then be found by solving the following spectral problem for the dominant eigenvalue \cite{touchette2017}:
\be
\cL_k r_k(\bx) = \lambda(k) r_k(\bx).
\label{spec1}
\ee
Since the tilted generator $\cL_k$ is not generally Hermitian, this spectral equation has to be considered in conjunction with the adjoint equation 
\be
\cL_k^{\dagger} l_k(\bx) = \lambda(k) l_k(\bx),
\label{spec2}
\ee
where $\cL_k^{\dagger}$ is the adjoint of $\cL_k$ and $l_k$ is the eigenfunction of $\cL_k^{\dagger}$ associated with its dominant eigenvalue \cite{touchette2017}. For convenience, we take these eigenfunctions to satisfy the normalization conditions
\be
\int_{\reals^n} r_k(\bx) l_k(\bx) d\bx = 1
\label{spec3}
\ee
and 
\be
\int_{\reals^n} l_k(\bx) d\bx = 1. 
\label{spec4}
\ee
The problem of obtaining the rate function is therefore reduced to the problem of solving the FK equation or solving a particular spectral problem for the process and observable considered.

\subsection{Effective process}

The spectral problem \eqref{spec1} determines not only the SCGF and, in turn, the rate function characterising the likelihood of the fluctuations of $A_T$, but also provides a way to understand how these fluctuations arise in the long-time limit in terms of an effective process that describes the subset of trajectories leading to a given fluctuation $A_T=a$ \cite{chetrite2013,chetrite2014,chetrite2015,jack2015}. This effective process, which is also called the auxiliary, driven or fluctuation process, was studied extensively for jump processes \cite{simon2009,popkov2010,jack2010b,grandpre2018,das2019} and SDEs \cite{angeletti2015,tsobgni2016,tsobgni2018,buisson2020b,tizon2019,yan2022}. In the latter case, it takes the form of a modified diffusion $\bX _k(t)$ satisfying the SDE
\be
d\bX _k(t) = \bF_k(\bX _k(t)) dt + \sigma d\bW (t),
\label{effproc}
\ee
which has the same noise matrix $\sigma$ as that of the original process, but with the \textit{effective drift} $\bF_k$ given by 
\be
\bF_k (\bx)= \bF(\bx) + D\bsn  \ln r_k(\bx)
\label{effdrift1}
\ee
for the additive observables (here, linear and quadratic) and
\be
\bF_k(\bx) = \bF(\bx) + D[k \obsm\bx+\bsn  \ln r_k(\bx)]
\label{effdrift2}
\ee
in the case of linear-current observables \cite{chetrite2014}. Here, $\bF(\bx)=-M\bx$ is again the original drift of the linear SDE, while $r_k(\bx)$ is the eigenfunction related to the dominant eigenvalue and SCGF $\lambda(k)$. Moreover, the value $k$ is set for a given fluctuation $A_T=a$ to $k_a$,  via the duality relation \eqref{duality}, which plays a role analogous to the temperature-energy relation in equilibrium statistical mechanics \cite{chetrite2014}.

The effective process or effective SDE is also ergodic \cite{chetrite2014} and, therefore, has a unique stationary density, known to be given by
\be
p^*_k(\bx) = r_k(\bx)l_k(\bx),
\label{pklkrk}
\ee
and a stationary current, given in general by
\be
\bJ^*_k(\bx) = \bF_k(\bx) p_k^*(\bx) - \frac{1}{2}D \bsn p_k^*(\bx).
\label{effstatcur1}
\ee
We study these modifications of the density and current, as well as the effective SDE supporting them, in the next sections for the three observables of interest. Since the effective process is ergodic, it also has a stationary value of $A_T$, denoted in the remaining by $a^*_k$, and given as in Eqs.~\eqref{meanadd}-\eqref{meancurr} by replacing $p^*$ and $\bJ^*$ with $p^*_k$ and $\bJ^*_k$, respectively. Mathematically, $a^*_k$ is also the inverse function of  $k_a$, following the duality relation \eqref{duality}, so that $a^*_k=\lambda'(k)$.

The effective SDE can be interpreted, as mentioned, as the SDE describing the subset of trajectories giving rise in the long-time limit to a fluctuation $A_T=a$, which has $p^*_{k_a}$ as its stationary density, $\bJ^*_{k_a}$ as its stationary current, and $a^*_{k_a}=a$ as its stationary and typical value of $A_T$ \cite{chetrite2014}. In general, $p^*_0 = p^*$ and $\bJ^*_0=\bJ^*$ for $k=0$, since the original process is not modified when observing its typical value $A_T=a^*$. Alternatively, it is known that the effective process can be interpreted as an optimal control process, whose drift minimizes a certain cost function in the long-time limit, related to the relative entropy \cite{chetrite2015}. From this point of view, the modified density and current can be seen as optimal density and current fluctuations leading to or creating a given fluctuation $A_T=a$. We further discuss these interpretations in Sec.~\ref{secother} and refer to the original works \cite{chetrite2013,chetrite2014,chetrite2015,jack2010b,jack2015} on the effective process for more details.

\section{Main results}

We derive in this section the exact generating function of $A_T$ for the three classes of observables defined before by solving the FK equation, and obtain from the result their SCGF and rate function by investigating the long-time limit of the generating function. We also obtain explicit expressions for the dominant eigenfunction $r_k$, which allows us to study the effective SDE, providing us with a clear understanding of how fluctuations of these observables arise from modified forces, densities, and currents in linear diffusions. To be concise, we provide only the final results for the various functions considered, which can be checked by direct substitution into the FK equation or the spectral equations. For more details about the derivation of these solutions, which follow by discretizing and iteratively solving the FK equation in time, we refer to \cite{buisson2022}.

\subsection{Linear additive observables}

We begin our analysis with the linear additive observable $A_T$, defined in \eqref{linadd}, which involves the vector $\obsv $ in the linear contraction with the state $\bX (t)$ of the linear SDE \eqref{SDE}. For this observable, the generating function $G_k(\bx,t)$ satisfies the FK equation \eqref{FK} with the tilted generator
\be
\cL_k = -\dprod{M\bx}{\bsn}  + \frac{1}{2} \dprod{\bsn}{ D \bsn}  + k  \dprod{\obsv}{\bx},
\label{tiltedgen1}
\ee
which is solved, for the initial condition $G_k(\bx,0)=1$, by
\be
G_k(\bx,t) = e^{\dprod{\bv_k(t)}{ \bx}} e^{\frac{1}{2} \int_0^t  \dprod{\bv_k(s)}{ D \bv_k(s)} ds},
\label{GKlinear}
\ee
where $\bv_k(t)$ is a vector in $\reals^n$ satisfying the differential equation 
\be
\frac{d \bv_k(t)}{dt} = k\obsv  - M^\transp\bv_k(t)
\label{differential1}
\ee
with initial condition $\bv_k(0) = \bs{0}$. 

This gives the exact generating function of $A_T$ at all times $t\geq 0$. To extract the SCGF from this result, we note that (\ref{differential1}) has a stationary solution $\bv^*_k$ given explicitly by
\be
\bv^*_k = k \big(M^\transp\big)^{-1} \obsv ,
\label{wkstat}
\ee
which is an attractive fixed point for all $k\in\reals$, since $M$ is assumed to be positive definite. As a result, we obtain from the definition (\ref{SCGF}) of the SCGF,
\be
\lambda(k) =\frac{1}{2}  \dprod{\bv^*_k}{ D \bv^*_k}
\label{SCGFlin}
\ee
or, more explicitly,
\be
\lambda(k) = \frac{k^2}{2} \dprod{\big(M^\transp\big)^{-1}\obsv} {D\big(M^\transp\big)^{-1} \obsv}  .
\label{SCGFlin2}
\ee
The fact that the result is quadratic in $k$ means that the fluctuations of $A_T$ are Gaussian, as expected for linear integrals of Gaussian processes, with zero asymptotic mean and asymptotic variance
\be
\lambda''(0) = \dprod{\big(M^\transp\big)^{-1}\obsv}{D\big(M^\transp\big)^{-1} \obsv}.
\ee
This can be seen more explicitly by taking the Legendre transform of $\lambda(k)$, which yields the quadratic rate function
\be
I(a) = \frac{a^2}{2 \dprod{\big(M^\transp\big)^{-1}\obsv} {D\big(M^\transp\big)^{-1} \obsv} }.
\ee

To understand physically how these Gaussian fluctuations arise, we note that $G_k(\bx,t)$ is known \cite{chetrite2014} to scale in the long-time limit according to
\be
G_k(\bx,t)\sim r_k(\bx) e^{t\lambda(k)},
\label{genscale}
\ee
so we can write directly
\be
r_k(\bx) = e^{\dprod{\bv^*_k}{\bx}}.
\ee
It can be verified that this function satisfies the spectral equation (\ref{spec1}) for $\cL_k$ as given in \eqref{tiltedgen1} and $\lambda(k)$ as given in (\ref{SCGFlin}), so it is indeed the dominant eigenfunction of $\cL_k$. Consequently, we find from \eqref{effdrift1} that the modified drift of the effective process is 
\be
\bF_k(\bx) = -M (\bx -  \bx^*_k),
\ee
where $\bx^*_k = M^{-1} D \bv^*_k$. Thus, the effective process is also a linear process with the same drift matrix $M$ as the original process, but with a fixed point in the drift pushed from the origin $\bx = \bs{0}$ to $\bx^*_k$ to create the fluctuation $A_T=a^*_k$. Its stationary density is therefore simply a translation of the stationary density of the original process, $p_k^*(\bx) = p^*(\bx-\bx^*_k)$, which is consistent with 
\be
a^*_k= \int_{\reals^n} \dprod{\obsv}{\bx}\, p^*_k(\bx)d\bx = \dprod{\obsv}{\bx^*_k}. 
\ee
Similarly, for the current we find
\be
\bJ^*_k(\bx) = H(\bx - \bx ^*_k) p_k^*(\bx) = \bJ^* (\bx-\bx^*_k).
\ee
These results confirm previous studies considering specific linear processes and linear observables \cite{coghi2021}, including the one-dimensional Ornstein--Uhlenbeck process \cite{chetrite2014}, and also confirm that the reversibility of the original SDE is not modified at the level of fluctuations \cite{chetrite2014}, since they only translate the current in space.

\subsection{Quadratic observables}

For the quadratic observable defined in \eqref{quadadd}, the tilted generator governing the evolution of the generating function has the form
\be
\cL_k = -\dprod{M\bx}{\bsn}  + \frac{1}{2} \dprod{\bsn}{D \bsn}  + k \dprod{\bx}{\obsQ \bx} .
\ee
and admits the following solution:
\be 
G_k(\bx,t) = e^{\dprod{\bx}{\ricmat_k(t) \bx}} e^{\int_0^t \Tr(D\ricmat_k(s)) ds},
\label{Gkquad}
\ee
where $\ricmat_k(t)$ is now a symmetric $n\times n$ matrix satisfying the differential Riccati equation 
\be
\frac{d\ricmat_k(t)}{dt} = 2 \ricmat_k(t) D \ricmat_k(t) - M^\transp \ricmat_k(t) - \ricmat_k(t)M + k\obsQ
\label{Akquaddiff}
\ee
with initial condition $\ricmat_k(0) = 0$. This can be obtained, as mentioned, by discretizing and iteratively solving the FK equation in time \cite{buisson2022}.

As before, the SCGF is determined by the stationary solution $\ricmat^*_k$ of this equation satisfying the algebraic Riccati equation 
\be
2\ricmat^*_k D \ricmat^*_k - M^\transp \ricmat^*_k - \ricmat^*_k M + k\obsQ = 0.
\label{Akquadricc}
\ee
In general this equation has multiple possible solutions; the correct one is found by requiring $\ricmat^*_{0} = 0$, since $G_{0}(\bx,t) = 1$ for all $\bx$ and $t$. Provided that this solution is a stationary solution of \eqref{Akquaddiff}, then the generating function scales in the long-time limit according to 
\be
G_k(\bx,t) \sim e^{\dprod{\bx}{\ricmat^*_k \bx}} e^{t \Tr(D\ricmat^*_k)},
\label{quadgenlim}
\ee
so that the SCGF is found to be
\begin{equation}
\lambda(k) = \Tr(D\ricmat^*_k).
\label{quadscgf}
\end{equation}

A similar result was found independently by Monthus and Mazzolo \cite{monthus2022} using a more complicated path integral approach. There are many results also in mathematics on the SCGF of quadratic observables of Gaussian processes \cite{benitz1990,bercu1997,bryc1997,gamboa1999,landais1999,bercu2000,bercu2002}, but most are expressed in terms of the spectral density of these processes. It is an open problem to establish an equivalence between these results and the trace result above involving the Riccati matrix. 

From the expression of the SCGF, we obtain the rate function $I(a)$ by Legendre transform. The result is not explicit, since $\ricmat^*_k$ must now be found by solving (\ref{Akquadricc}). However,  it can be checked from this equation that the asymptotic mean of $A_T$, which corresponds to the zero of $I(a)$, is
\be
a^*=\lambda'(0)= \Tr(\obsQ\covm),
\label{quadmean}
\ee
consistent with \eqref{meanquad}. Moreover, the asymptotic variance is
\be
\lambda''(0) = 4 \Tr(\covm \obsQ\covm {\ricmat^*_0}' ),
\ee
where ${\ricmat^*_0}'$ is the derivative of $\ricmat^*_k$ with respect to $k$ evaluated at $k=0$, which satisfies yet another Lyapunov equation
\be
M^\transp {\ricmat^*_0}' +{\ricmat^*_0}' M = \obsQ.
\label{ricc1}
\ee
The full derivation of these results can be found in \cite{buisson2022}. The variance result is important, as it gives the variance of the small Gaussian fluctuations of $A_T$ around $a^*$, determined by expanding $I(a)$ to second order around $a^*$. Large fluctuations of $A_T$ away from this value are generally not Gaussian, since $I(a)$ is generally not quadratic for quadratic observables, as shown in Sec.~\ref{secapps}.

To understand how these small and large fluctuations are created, we note again the scaling in \eqref{genscale} to infer from \eqref{quadgenlim}:
\be
r_k(\bx) = e^{\dprod{\bx}{\ricmat^*_k\bx}}.
\label{quadrk}
\ee
It can be checked again that this solves the spectral equation (\ref{spec1}) with the eigenvalue given in (\ref{quadscgf}). From \eqref{effdrift1}, we then find 
\be
\bF_k(\bx) = -M_k \bx,
\label{effdriftquad}
\ee
where
\be
M_k = M - 2D\ricmat^*_k.
\label{mkquad}
\ee
Hence we see that the effective process associated with quadratic observables is still a linear diffusion, but now involves a modified drift matrix, leading to the following stationary density:
\be
p^*_k(\bx) = \sqrt{\frac{1}{(2\pi)^n \det \covm_k}}\exp\left(-\frac{1}{2} \dprod{\bx}{\covm_k^{-1} \bx} \right),
\label{rhoquad}
\ee
where $\covm_k$ is the modified covariance matrix satisfying the Lyapunov equation
\be
D = M_k \covm_k + \covm_k M_k^\transp.
\label{ricccovquad}
\ee
Moreover, the associated current is
\be
\bJ^*_k(\bx) = H_k\bx p_k^*(\bx),
\label{Jquad}
\ee
where the matrix $H_k$ is obtained from \eqref{currmat} by replacing $\covm$ and $M$ by $\covm_k$ and $M_k$, respectively. 

It is interesting to note from these results that, for a reversible SDE with $M$ symmetric and $D\propto\id$, the effective process remains reversible, so that $\bJ^*_k=\bs{0}$ if $\bJ^*=\bs{0}$. In this case, only the density $p^*$ is modified to $p^*_k$ to create fluctuations of $A_T$. This can be checked from \eqref{Jquad}, but follows more easily by noting from \eqref{effdrift1} that $\bF_k$ is gradient if $\bF$ itself is gradient when $D\propto\id$. On the other hand, for an irreversible SDE, the density \emph{and} the current are generally modified to accommodate fluctuations, as predicted by \eqref{rhoquad} and \eqref{Jquad}, so the irreversible properties of the effective SDE can differ in this case from those of the original SDE, as illustrated in Sec.~\ref{secapps}.

To close, we should note that the results for $p^*_k$ and $\bJ^*_k$ above hold if the effective process is ergodic, that is, if $M_k$ is positive definite. Although obvious, this is an important remark because it provides us with a criterion for determining whether $\lambda(k)$ exists for a given $k$, which is easier to check than the criterion mentioned earlier about the existence of stationary solutions of the time-dependent Riccati equation \eqref{Akquaddiff}. If $M_k$ is not positive definite for a given $k$, then the Lyapunov equation (\ref{ricccovquad}) does not have a positive definite solution $C_k$ and, as such, the eigenfunction $r_k$, formally expressed in (\ref{quadrk}), does not constitute a valid eigenfunction of the spectral problem associated with the SCGF, which implies that the SCGF itself does not exist.

\subsection{Current-type observables}

We conclude by considering linear current-type observables, as defined in (\ref{lincurobs}), which involve an $n \times n $ matrix $\obsm$. We first address the case where $\obsm$ is purely antisymmetric so that $\obsm = -\obsm^\transp $. The case where $\obsm$ also has a non-zero symmetric part is more involved and is therefore treated separately after. 

\begin{widetext}
\subsubsection{Antisymmetric $\obsm$}

For the observable (\ref{lincurobs}), with $\obsm$ assumed to be purely antisymmetric, the associated tilted generator $\cL_k$ is given by 
\be
\cL_k = -k \dprod{M\bx}{\obsm\bx} + \dprod{(-M + kD\obsm)\bx}{\bsn}  + \frac{1}{2} \dprod{\bsn}{D\bsn} + \frac{k^2}{2}  \dprod{\obsm\bx}{D\obsm\bx} .
\ee
This can be written in a slightly more convenient form as 
\be
\cL_k = -\frac{k}{2}\dprod{\bx}{(M^\transp \obsm - \obsm M)\bx} +  \dprod{(-M + kD\obsm)\bx}{\bsn} + \frac{1}{2} \dprod{\bsn}{D\bsn} + \frac{k^2}{2}  \dprod{\bx}{\obsm^\transp D\obsm\bx},
\label{tiltedcurrentoperator}
\ee
given that 
\be
\dprod{M\bx}{\obsm\bx} =  \dprod{\bx}{M^\transp \obsm\bx} = \frac{1}{2}\dprod{\bx}{ (M^\transp \obsm +  \obsm^\transp M)\bx} = \dprod{\bx}{ (M^\transp \obsm - \obsm M)\bx},
\ee
where we have used the antisymmetry of $\obsm$ in the last equality. 

The solution of the FK equation with the tilted generator (\ref{tiltedcurrentoperator}) is the same as that found in \eqref{Gkquad} for quadratic additive observables, except that the differential Riccati equation satisfied by $\ricmat_k(t)$ is now 
\be
\frac{d\ricmat_k(t)}{dt} = \frac{k^2}{2} \obsm^\transp D\obsm - \frac{k}{2}(M^\transp \obsm-\obsm M) + (-M + kD\obsm)^\transp \ricmat_k(t) + \ricmat_k(t)(-M + kD\obsm) + 2\ricmat_k(t) D\ricmat_k(t),
\label{riccatidiffcur}
\ee
with initial condition $\ricmat_k(0) = 0$. A similar equation was obtained using path-integral methods for a particular type of linear current-type observable, namely, the nonequilibrium work, by Kwon, Noh and Park~\cite{kwon2011}, who then obtained large deviation results for this observable via numerical integration. Here, we obtain the SCGF and rate function directly by considering the stationary solution $\ricmat^*_k$ of the Riccati equation, which now satisfies the algebraic Riccati equation
\be
\frac{k^2}{2} \obsm^\transp D\obsm - \frac{k}{2}(M^\transp \obsm-\obsm M) + (-M + kD\obsm)^\transp \ricmat^*_k + \ricmat^*_k(-M + kD\obsm) + 2\ricmat^*_k D\ricmat^*_k = 0
\label{ricccur} 
\ee
with $\ricmat^*_0 = 0$. Assuming, as before, that the correct solution of this equation is a stationary solution of \eqref{riccatidiffcur}, we then recover the same expression of the SCGF as for quadratic observables, namely, 
\be
\lambda(k) = \Tr (D\ricmat^*_k),
\label{curscgf}
\ee
from which we obtain the rate function $I(a)$ by Legendre transform. The results again are not explicit, but rely nevertheless on the solution of \eqref{ricccur}. From this equation, it can also be checked as before that the asymptotic mean of $A_T$ is the one found in \eqref{meancurr}, while the asymptotic variance, characterizing the Gaussian regime of fluctuations near $a^*$, is
\be
 \lambda''(0) = \Tr\left[\covm \obsm^\transp D\obsm + 2 \covm(\obsm M - M^\transp \obsm ) \covm {\ricmat^*_0}' + 2 \covm (\obsm^\transp D {\ricmat^*_0}' +  {\ricmat^*_0}' D\obsm)\right],
 \label{curvar}
\ee
where ${\ricmat^*_0}'$ now satisfies the Lyapunov equation
\be
{\ricmat^*_0}' M + M^\transp {\ricmat^*_0}' = \frac{1}{2}\left(\obsm M - M^\transp \obsm \right).
\label{bigricc1}
\ee
We show in the application section that these equations can be solved exactly in non-trivial cases.

\end{widetext}

Since the generating function has the same form as that obtained for quadratic observables, the eigenvector $r_k$ has also the same form as that shown in \eqref{quadrk}, which means that the effective process is again a linear SDE with a drift matrix entering in \eqref{effdriftquad} now given by
\be
M_k = M - 2D\ricmat^*_k - k D \obsm.
\label{mkcur}
\ee
As before, for those $k$ for which $M_k$ is positive definite, the effective process is ergodic and large deviations exist. In this case, the stationary density $p^*_k$ has the same form as (\ref{rhoquad}), using $M_k$ as above in the Lyapunov equation for the covariance matrix $\covm_k$. Similarly, the modified current $\bJ^*_k$ is given as in \eqref{Jquad}, using the appropriate $\covm_k$ and $M_k$ for the current observable.

Despite the fact that the effective SDEs associated with the quadratic and current-type observables have the same linear form, they have different reversibility properties coming from their different $M_k$. In particular, for current-type observables, the effective process is in general irreversible even if the original process is reversible, since a current has to be produced to sustain a non-zero fluctuation of $A_T$. This follows by noting that the effective drift for this observable, shown in \eqref{effdrift2}, has an added part involving $\obsm$, which is non-gradient when $D\obsm$ is not symmetric. In a more obvious way, we also know that a fluctuation of $A_T$ in the original process is realized as the typical value
\be
a^*_k = \int_{\reals^n} \dprod{\obsm\bx}{\bJ^*_k}\, p^*_k(\bx)d\bx
\ee
in the effective process, so that $\bJ^*_k\neq 0$ if $a^*_k\neq 0$. The same relation applies for irreversible SDEs and implies for those that the current is modified by fluctuations. In particular, for $a^*_k=0$, we have $\bJ^*_k=\bs{0}$, so an irreversible process can behave as a reversible process when conditioned on observing the fluctuation $A_T=0$. An example of this unusual fluctuation behavior is discussed in Sec.~\ref{secapps}.

\subsubsection{General $\obsm$}

We now address the case where the matrix $\obsm$ has a non-zero symmetric component. To this end, we decompose this matrix as $\obsm = \obsm^+ + \obsm^-$ in terms of its symmetric and antisymmetric parts
\be
\obsm^{\pm} = \frac{\obsm \pm \obsm^\transp}{2},
\ee 
so as to express the observable similarly as $A_T = A_T^++A_T^-$, where  
\be
A_T^{\pm} = \frac{1}{T} \int_0^T \obsm^{\pm} \bX (t)\circ d\bX (t).
\ee
We have already discussed the antisymmetric part $A_T^-$ before. As for the symmetric part, we can integrate it directly to obtain
\be
 A_T^+ = \frac{1}{2T}  \dprod{\bX (T)}{\obsm^+ \bX (T)} -\frac{1}{2T} \dprod{\bX (0)}{\obsm^+\bX (0)},
 \label{eqboundary}
\ee
since the Stratonovich convention used in the definition of the observable preserves the standard rules of calculus. Thus, this part only adds boundary terms to $A_T^-$, which can contribute to the large deviations of $A_T$, surprisingly, even though they are not extensive in time, because they can limit the range of values of $k$ for which $\lambda(k)$ exists. 

This effect was described for particular observables in recent studies \cite{kundu2011b,sabhapandit2011,sabhapandit2012}, and can be understood by expressing the generating function as
\be
G_k(\bx,t) = \int_{\reals^n}d\by \, G_k(\bx,\by,t),
\ee
where 
\be
G_k(\bx,\by,t) = \E[\delta (\bX (t)-\by) e^{tkA_t}|\bX(0)=\bx]
\ee
is the generating function of $A_T$ in which both $\bX (0)$ and $\bX (t)$ are fixed. Considering the decomposition of $A_T$ above, we then have
\be
G_k(\bx,t) = e^{- \frac{k}{2} \dprod{\bx}{\obsm^+\bx}} \int_{\reals^n} d\by\, e^{\frac{k}{2} \dprod{\by}{\obsm^+\by}}\, G^-_k(\bx,\by,t),
\label{genint}
\ee
$G^-_k(\bx,\by,t)$ being the generating function of $A_T^-$ with fixed initial and terminal states. 

In the long-time limit, it is known \cite{chetrite2014} that this generating function scales, similarly to \eqref{genscale}, according to 
\be
G_k^-(\bx,\by,t)\sim e^{t\lambda^-(k)} r_k^-(\bx)l^-_k(\by),
\ee 
where $\lambda^-(k)$ is the dominant eigenvalue and SCGF of $A_T^-$ with eigenfunctions $r_k^-$ and $l^-_k$. This eigenvalue was already obtained in \eqref{curscgf}, while $r^-_k$ was found in \eqref{quadrk} with $\ricmat^*_k$ satisfying the algebraic Riccati equation \eqref{ricccur} for $\obsm^-$. As for $l^-_k$, we can find it using \eqref{rhoquad} with $M_k$ given as in \eqref{mkcur}, leading to
\be
G_k^-(\bx,\by,t) \sim e^{t\lambda^-(k)} e^{-\dprod{\by}{\ricmat^*_k\by} -\frac{1}{2}\dprod{\by}{\covm_k^{-1}\by} + \dprod{\bx}{\ricmat^*_k \bx}},
\ee
up to a multiplicative constant, and thus to
\be
G_k(\bx,t) \sim e^{t\lambda^-(k)} e^{\dprod{\bx}{\left(\ricmat^*_k - \frac{k}{2}\obsm^+\right)\bx }}\int_{\mathbb{R}^n} d\by \, e^{-\frac{1}{2}\dprod{ \by}{ \cB_k\by}},
\label{longtimegen}
\ee
where 
\be
\cB_k = \covm_k^{-1} + 2\ricmat^*_k - k\obsm^+.
\label{posdef}
\ee
In this last expression, both $\covm_k$ and $\ricmat^*_k$ are associated with $\obsm^-$, and are thus obtained by following our previous results for antisymmetric current observables. 

The difference now for general current observables is that, for $G_k(\bx,t)$ to exist, the integral over $\by$ above needs to be convergent, which holds when $\cB_k$ is positive definite. In this case, we obtain $\lambda(k) = \lambda^-(k)$, assuming that $\lambda^-(k)$ itself exists. If $\cB_k$ is not positive definite, then $\lambda(k)= \infty$, so the domain where the SCGF exists is effectively cut or limited by $\obsm^+$ \cite{kundu2011b,sabhapandit2011,sabhapandit2012}. 

To express this result more precisely, let us denote by $\cK^-$ the interval of $k$ values for which the SCGF $\lambda^-(k)$ associated with $\obsm^-$ exists, which, we recall, is determined by requiring that $\covm_k$ is positive definite. Moreover, let $\cK^+$ denote the interval of values for which $\cB_k$ is positive definite. Then
\be
\lambda(k) = 
\begin{cases}
\lambda^-(k) & k\in \cK^-\cap\cK^+ \\ 
\infty & \textrm{otherwise}. 
\end{cases}
\ee

In general, the intersection of $\cK^-$ and $\cK^+$ defines a specific value of $k$ beyond which the SCGF ceases to exist. For concreteness, we can take this cut-off value to be positive, denoting it by $k_{\max}$, to rewrite the SCGF as
\be
\lambda(k) = 
\begin{cases}
\lambda^-(k) & k<k_{\max} \\ 
\infty & k \geq k_{\max}.
\end{cases}
\ee
From the properties of Legendre transforms, it is known that the existence of the cut-off $k_{\max}$ has the effect of creating a linear branch in the rate function $I(a)$ beyond a point $\ta$ given by the left derivative of $\lambda^-(k)$ at $k_{\max}$ (see \cite[Ex.~3.3]{touchette2009}). As a result, the rate function of $A_T$ can be written as
\be
I(a) = \begin{cases}
I^-(a) & a < \ta \\ 
k_{\max}a - \lambda^-(k_{\max}) & a \geq \ta,
\end{cases}    
\label{ratelinear}
\ee
where $I^-(a)$ is the rate function associated with $A_T^-$. Therefore, we see that the fluctuations of $A_T$ \emph{below} $\ta$ are determined by the fluctuations of the antisymmetric (time-extensive) part $A_T^-$, with the boundary term $A_T^+$ playing no role, whereas the fluctuations of $A_T$ \emph{above} $\ta$ are determined by $A_T^+$ and, more specifically, by the term in \eqref{eqboundary} involving $\bX (T)$, since $\bX (0)$ is fixed here to $\bx$. If the initial condition is chosen instead according to a probability density $p(\bx,0)$, then there is usually another cut-off, $k_{\min}<0$, coming from the integration of $\bX (0)$ over that density \cite{kim2014}. In this case, $I(a)$ generally has two linear branches, instead of one, related to the fluctuations of $A_T^+$ coming from the initial and terminal boundary terms. This type of rate function has been studied before, in particular, in the context of the so-called extended fluctuation relation \cite{zon2003}.

To close this section, we note that because the effective process is based on $r_k$, it is defined only in the domain of the SCGF, here $k<k_{\max}$, describing the fluctuations of $A_T$ dominated by those of $A_T^-$. In that region, the result in \eqref{longtimegen} implies
\be
r_k(\bx)=e^{\dprod{\bx}{\left(\ricmat^*_k - \frac{k}{2}\obsm^+\right)\bx}},
\ee
so that
\be
\bF_k(\bx) = -M\bx + kD\obsm\bx + D\left(2\ricmat^*_k - k\obsm^+\right)\bx.
\ee
However, since $\obsm = \obsm^+ + \obsm^-$, this becomes 
\be
\bF_k(\bx) = -(M + 2D\ricmat^*_k +kD\obsm^-) \bx,
\label{effdriftgenlin}
\ee
which is exactly the effective drift associated with $\obsm^-$, as given by \eqref{mkcur}, confirming that the boundary terms in the observable play no role. For the regime of fluctuations of $A_T$ dominated by these terms, it is not known what the effective process is or even if such a process exists \cite{tsobgni2018}. At the very least, it cannot be defined from spectral elements, since those elements do not exist outside the domain of the SCGF.

\section{Other approaches}
\label{secother}

It is known in large deviation theory that the SCGF can be obtained from two other approaches related to control theory and optimization \cite{chetrite2015}. We briefly discuss them here to complete our results and to establish a link with the classical Gaussian control problem. For simplicity, we consider only the case of additive linear observables. Similar results apply for the two other observables.

The first approach is based on the idea of modifying the drift of the original SDE to obtain a new SDE with drift $\tbF$, assumed also to be ergodic. By considering all such modifications, it is known \cite{chetrite2015} that the SCGF of $A_T$ with respect to the original SDE can be expressed in a variational way as
\be
\lambda(k) = \lim_{T\ra\infty}\max_{\tbF} \E[k A_T -R_T],
\label{opt1}
\ee
where $\E[\cdot]$ now denotes the expectation with respect to the modified SDE and
\begin{widetext}
\be
R_T = \frac{1}{2T}\int_0^T \dprod{\bF(\bX(t))-\tbF(\bX(t))}{D^{-1}[\bF(\bX(t))-\tbF(\bX(t))]} dt
\ee
is a time-averaged ``distance'' between the modified and original SDEs. Equivalently, since the modified SDE is assumed to be ergodic, we can replace the long-time expectation by an expectation involving the stationary density of this process, denoted by $\tp$, so as to write
\be
\lambda(k) =  \max_{\tbF} \{k A[\tp]- R[\tp,\tbF]\},
\label{opt2}
\ee
where
\be
A[\tp] = \int_{\reals^n} \dprod{\obsv}{\bx}\, \tp(\bx)\, d\bx
\ee
is, similarly to \eqref{meanadd}, the typical value of $A_T$ in the modified SDE and
\be
R[\tp,\tbF] = \int_{\reals^n} \dprod{\bF(\bx)-\tbF(\bx)}{D^{-1}(\bF(\bx)-\tbF(\bx))}\, \tp(\bx)\, d\bx
\ee
is the typical distance.
\end{widetext}

The two maximization problems in \eqref{opt1} and \eqref{opt2} have a natural interpretation in terms of a controlled SDE whose drift is modified so as to maximize the cost or loss function $K_T = \E[kA_T-R_T]$ \cite{stengel1994,schulz2006,bechhoefer2021}. The control is applied over an infinite time horizon, leading to an ergodic process that realizes the SCGF as the maximal cost. From recent works \cite{chetrite2014,chetrite2015,jack2015}, it is known that this optimal control process is the effective process described earlier with drift $\bF_k$ and stationary density $p^*_k$, so we can write in fact
\be
\lambda(k) = kA[p^*_k] - R[p^*_k, \bF_k].
\label{opt2sol}
\ee
The results of the previous section therefore predict that the optimal SDE that maximizes the cost $K_T$ in the long-time limit is a linear SDE characterized by a modified fixed-point or a modified drift matrix $M_k$ satisfying an algebraic Riccati equation.


From a control perspective, these results can be derived by assuming that the control drift $\tbF$ is linear in the state. In this case, the cost $K_T$ has a linear part and a quadratic part in $\bx$, which has been studied extensively as the linear-quadratic-Gaussian (LQG) control problem \cite{stengel1994,schulz2006,bechhoefer2021}. It can be checked that the well-known Riccati equation associated with this problem  recovers the results found here for the three observables considered, with the following differences:
\begin{itemize}
\item The LQG problem is formulated by minimizing linear-quadratic cost functions over the class of ergodic controls that are linear in the control inputs, leading to an optimal controller that is linear in $\bx$. Here, we make no such linearity assumption; the linearity of the optimal controller follows from the spectral solution giving $\lambda(k)$ and $r_k(\bx)$. 

\item The quadratic part of the cost function in LQG control is assumed to be positive definite to guarantee that the minimization problem has a solution. In our case, that part is not necessarily positive definite, depending on the observable and $k$ value considered, because the SCGF is not necessarily positive. However, the minimization has a solution if the SCGF exists.

\item For current-type observables, the functional $A[\tp]$ involves the stationary current $\tbJ$ of the controlled SDE, similarly to \eqref{meancurr}, instead of its stationary density, giving rise to a control cost involving the density and current of the control process or, equivalently, its state and increments \cite{bierkens2014,chernyak2014,bierkens2013}, which generalizes the classical LQG control problem.
\end{itemize}

The SCGF can be obtained in a slightly different way by noting, as done earlier, that the drift of an ergodic SDE is uniquely determined by its stationary density and current, so the minimization in \eqref{opt2} over $\tbF$ can be re-expressed as a minimization over densities $\tp$ that are normalized in $\reals^n$ and currents $\tbJ$ that satisfy the stationary condition \eqref{current}. This change of variables has the effect of transforming the distance $R[\tp,\tbF]$ to
\begin{widetext}
\be
I[\tp,\tbJ] = \frac{1}{2}\int_{\reals^n} \dprod{(\tbJ(\bx)-\bJ^*_{\tp}(\bx))}{(\tp(\bx) D)^{-1}(\tbJ(\bx)-\bJ^*_{\tp}(\bx))}\, d\bx,
\ee
where $\bJ^*_{\tp}$ is an ``instantaneous'' current obtained from \eqref{current} by replacing $p^*$ with $\tp$ \cite{chetrite2015}. As a result, the minimization in \eqref{opt2} giving the SCGF becomes
\be
\lambda(k) = \max_{\tp,\tbJ} \{k A[\tp] - I[\tp,\tbJ]\}.
\label{opt3}
\ee
\end{widetext}

This result plays a special role in large deviation theory, as the functional $I[\tp, \tbJ]$ has the interpretation of a rate function, characterizing the probability that the original SDE with drift $\bF$ gives rise to a density fluctuation $\tp$ away from its stationary density $p^*$ concurrently with a current fluctuation $\tbJ$ away from its stationary current $\bJ^*$ \cite{maes2008a,maes2008b,barato2015,hoppenau2016}. From this point of view, the maximization in \eqref{opt3} can be seen as a Lagrange version of the problem of finding the most likely density and current fluctuations that give rise to a fluctuation $A_T=a$ of the observable, with $k$ playing the role of the Lagrange parameter \cite{chetrite2015}. Many works have appeared recently on this level of fluctuations \cite{maes2008a,maes2008b,barato2015,hoppenau2016}, known technically as the level 2.5 of large deviations, so we refer to them for more details.

To be consistent with the solution \eqref{opt2sol}, the optimal density and current fluctuations that are most likely to appear must correspond to the stationary density and current of the effective process, so we also have
\be
\lambda(k) = k A[p^*_k] - I[p^*_k,\bJ^*_k].
\ee
It can be checked that this result, as well as the one shown in \eqref{opt2sol}, agree with the explicit expressions that we have found in the previous section for $\lambda(k)$, $\bF_k$, $p^*_k$ and $\bJ^*_k$, which means that these expressions can be derived, in principle, by solving the ergodic control problem in \eqref{opt2} or the optimization problem in \eqref{opt3}. This also applies for quadratic and current-type observables. The only difference for the latter is that $A[\tp]$ is not a function of the density but of the current $\tbJ$, similarly to \eqref{meancurr}.

\section{Applications}
\label{secapps}

We illustrate our results in this section with two examples of SDEs in $\reals^2$, used in statistical physics as minimal models of nonequilibrium systems, focusing on quadratic observables and linear current-type observables, as the case of linear additive observables is trivial. Some of the SDEs and observables that we consider have been studied before \cite{monthus2022,chernyak2006,turitsyn2007,kwon2011,noh2013,noh2014}, using different methods, however, based on path integrals. We revisit them here to show how the SCGF and rate function can be obtained in a more direct way using our approach based on Riccati equations, and extend these results by describing how different fluctuation regimes arise physically via density and current fluctuations related to the effective process.

\subsection{Quadratic observable for transverse diffusions}

The first system that we consider is the normal or transverse diffusion in $\reals^2$, defined by the general linear SDE \eqref{SDE} with
\be
M = 
\begin{pmatrix} 
\gamma && \xi \\ 
-\xi && \gamma 
\end{pmatrix}
\ee
and $\sigma = \epsilon\id$ with $\gamma>0$, $\xi\in\reals$, and $\epsilon>0$. This process serves as a minimal model of nonequilibrium steady-state systems \cite{chernyak2006,turitsyn2007,noh2013,noh2014}, since the antisymmetric part of the drift involving the parameter $\xi$ creates a stationary current given by
\be
\bJ^*(\bx) = \xi \begin{pmatrix} -x_2 \\ x_1 \end{pmatrix}p^*(\bx),
\label{Jtrans}
\ee
which involves the Gaussian stationary density
\be
p^*(\bx) = \frac{\gamma}{\pi \epsilon^2} e^{-\gamma \| \bx\|^2/\epsilon^2},
\label{rhotrans}
\ee
so that $\covm=\epsilon^2/(2\gamma)\id$. For $\xi < 0$, the current circulates around the origin in a clockwise direction, whereas, for $\xi > 0$, it circulates in an anticlockwise direction. When $\xi=0$, the current vanishes, giving rise to an equilibrium system with a gradient drift, which has the same stationary density as the nonequilibrium system, interestingly, since $p^*$ does not depend on $\xi$.

The first observable that we study for this system is the time-averaged squared distance from the origin:
\be
A_T = \frac{1}{T} \int_0^T \|\bX (t)\|^2dt, 
\ee
which corresponds to the choice $\obsQ=\id$ in the general quadratic observable (\ref{quadadd}). For this observable, the differential Riccati equation (\ref{Akquaddiff}) can be solved exactly to obtain $\ricmat_k(t)$ and, in turn, $G_k(\bx,t)$, which have well-defined limits, giving the SCGF $\lambda(k)$ \cite{buisson2022}. Alternatively, we can solve the algebraic Riccati equation (\ref{Akquadricc}) to obtain the steady-state solution  $\ricmat^*_k$ directly, yielding in both cases the diagonal matrix $\ricmat^*_k = \ricval^*_k \id$ with
\be
\ricval^*_k= \frac{\gamma - \sqrt{\gamma^2 - 2k\epsilon^2}}{2\epsilon^2}
\label{akstat}
\ee
for $k\in (-\infty, \gamma^2/(2\epsilon^2))$. Consequently, we find from \eqref{quadscgf},
\be
\lambda(k) =2\epsilon^2 \ricval^*_k = \gamma - \sqrt{\gamma^2 - 2k\epsilon^2}
\label{transquadSCGF}
\ee
for the same range of $k$ values. Taking the Legendre transform, we then obtain the following rate function:
\be
I(a) = \frac{\gamma^2 a}{2\epsilon^2} + \frac{\epsilon^2}{2 a} - \gamma,\quad a>0.
\ee

\begin{figure*}[t]
\centering
\includegraphics{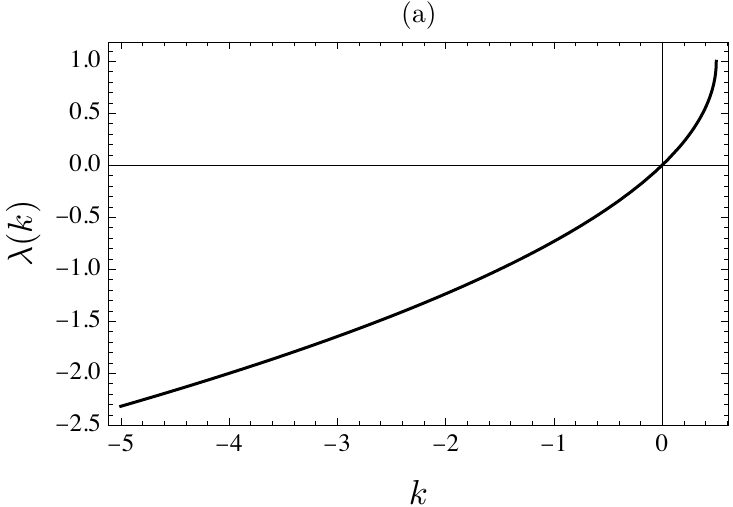}%
\hspace*{0.5in}
\includegraphics{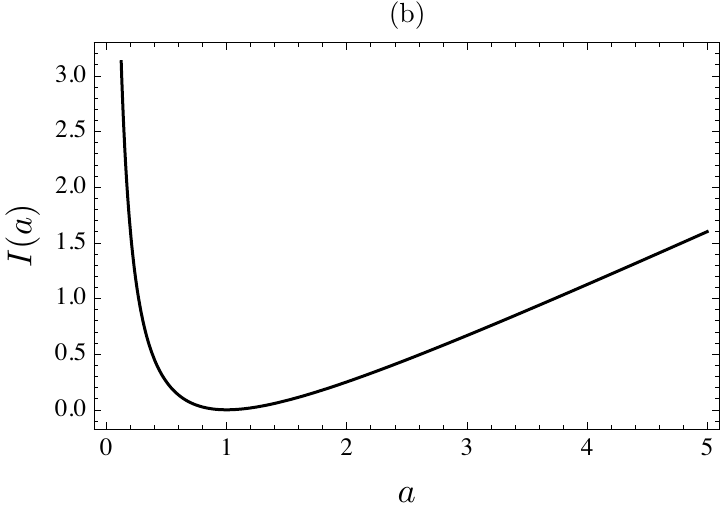}
\caption{(a) SCGF and (b) rate function of the squared norm of the transverse diffusion with $\gamma = 1,\xi=1$, and $\epsilon=1$.}
\label{fig:TransquadSCGFrate}
\end{figure*}

These two functions, plotted in Fig.~\ref{fig:TransquadSCGFrate}, are similar to those found for the one-dimensional Ornstein-Uhlenbeck process \cite{bryc1997}. The minimum of $I(a)$, giving the typical value of $A_T$, is $a^* = \epsilon^2/\gamma$, while the asymptotic variance is 
\be
\lambda''(0)=I''(a^*)^{-1} = \frac{\epsilon^4}{\gamma^3}.
\ee
This variance describes again the Gaussian fluctuations of $A_T$ in the vicinity of $a^*$. Away from this value, the fluctuations are non-Gaussian, as is clear from the form of $I(a)$. In fact, as $a\ra 0$, the term $\epsilon^2/(2a)$ dominates, so the right tail of $p(A_T=a)$ follows an inverse exponential distribution, while, for $a\ra\infty$, the term $\gamma^2 a/(2\epsilon^2)$ takes over, predicting an exponential distribution for the large values of $A_T$ generated by trajectories of the SDE venturing far away from the origin.

It is remarkable that both the SCGF and rate function are independent of the nonequilibrium parameter $\xi$. Intuitively, this can be understood by noting that $A_T$ is radially symmetric and, therefore, is not affected by the rotation of $\bX(t)$ around the origin. What matters is the distance of the trajectories of $\bX(t)$ from the origin, which is controlled by the diagonal (symmetric) part of the drift matrix $M$. Thus, small values of $A_T$ below $a^*$ must arise from trajectories that remain close to the origin, irrespective of the manner in which they rotate around this point, and should therefore be described by an effective drift that confines the process around the origin. Similarly, large fluctuations of $A_T$ above $a^*$ should arise from rare trajectories that are less confined around the origin but rotate freely around the origin as in the original process.

This is confirmed by calculating the effective drift matrix from \eqref{mkquad} to obtain
\be
M_k = \begin{pmatrix}\sqrt{\gamma^2 - 2k\epsilon^2} && \xi \\ -\xi && \sqrt{\gamma^2 - 2k\epsilon^2} \end{pmatrix}.
\label{Fktransquadform}
\ee
The diagonal part of this matrix is modified by $k$, resulting in an effective density with covariance
\be
\covm_k =  \frac{\epsilon^2}{2 \sqrt{\gamma^2 - 2k \epsilon^2}}\mathbb{I},
\ee
which is more or less confined around the origin, depending on the fluctuations considered. The antisymmetric part of the drift, on the other hand, remains the same, implying that the current is not modified in form. In fact, from \eqref{Jquad} we find
\be
\bJ^*_k(\bx) = \xi \begin{pmatrix}-x_2 \\ x_1 \end{pmatrix} p_k^*(\bx).
\ee
so the effective current differs from $J^*$ only to the extent that $p_k^*$ differs from $p^*$. The fluctuations of this observable are thus realized optimally by altering the density, with the only changes to the current resulting from those density modifications. In particular, if $\bJ^*=\bs{0}$, then $\bJ^*_k=\bs{0}$, so the reversibility of the original process is not changed for $\xi=0$ when looking at fluctuations. 

Of crucial importance for this to hold is the fact that the diffusion matrix $D$ is proportional to the identity and that the diagonal part of $M$ is proportional to the identity. If either or both of these properties are not satisfied, then $\ricmat^*_k$ can be non-diagonal, implying a non-trivial coupling of the density and current with $\xi$, even though $A_T$ is a density-type observable.

\subsection{Nonequilibrium work and entropy production for transverse diffusions}

The drift acting on an SDE can be seen as a force that performs work, which can be transformed in time into internal energy or dissipated as heat into the environment, depending on the physical system considered. Recently, these quantities have come to be studied as part of the stochastic thermodynamics (or stochastic energetics) formalism, which is concerned with extending the notions and laws of thermodynamics to stochastic processes \cite{sekimoto2010,seifert2012,peliti2021}. In this context, quantities such as work, heat and entropy take the form of time-integrated functionals of the system's state, which means that they are dynamical observables, and can be shown to satisfy conservation laws that generalize the first and second laws of thermodynamics. 

Two of the most important quantities in stochastic thermodynamics are the \emph{entropy production}, defined for the linear SDE \eqref{SDE} as
\be
\ep = -\frac{1}{T} \int_0^T 2 D^{-1} M\bX(t) \circ d\bX(t)
\label{entropyproduction}
\ee
and the \emph{nonequilibrium work}
\be
\work =-\frac{1}{T} \int_0^T 2 (D^{-1}M)^- \bX(t) \circ d\bX(t),
\label{neqwork}
\ee
which is the antisymmetric part of $\ep$ not related to a change of potential energy. The large deviations of these observables were studied by Noh \cite{noh2014} for transverse diffusions using path integral methods. We revisit them here to illustrate our simpler approach, and extend these results by discussing the properties of the effective process, which provides a physical way of understanding how large deviations arise in terms of modified drifts, densities and currents.

We begin by considering the nonequilibrium work, which is an antisymmetric current-type observable described for this SDE by the matrix 
\be
\obsm = 
\begin{pmatrix}
0 && -2\xi/\epsilon^2 \\ 
2\xi/\epsilon^2 && 0 
\end{pmatrix}.
\ee
Similarly to the quadratic observable, the SCGF of $\work$ can be obtained by either solving the time-dependent Riccati equation \eqref{riccatidiffcur} exactly, and by taking the long-time limit of the solution, or by solving the time-independent Riccati equation \eqref{ricccur}. The result in both cases is
\be
\lambda(k) = \gamma - \sqrt{\gamma^2 - 4k(1+k)\xi^2}
\label{scgftranswork}
\ee
for $k$ in the range
\be
\cK^-=\bigg(\frac{-\xi^2 - \sqrt{\gamma^2 \xi^2+ \xi^4}}{2\xi^2}, \frac{-\xi^2 + \sqrt{\gamma^2 \xi^2+ \xi^4}}{2\xi^2} \bigg).
\label{krange}
\ee

\begin{figure*}[t]
\centering
\includegraphics{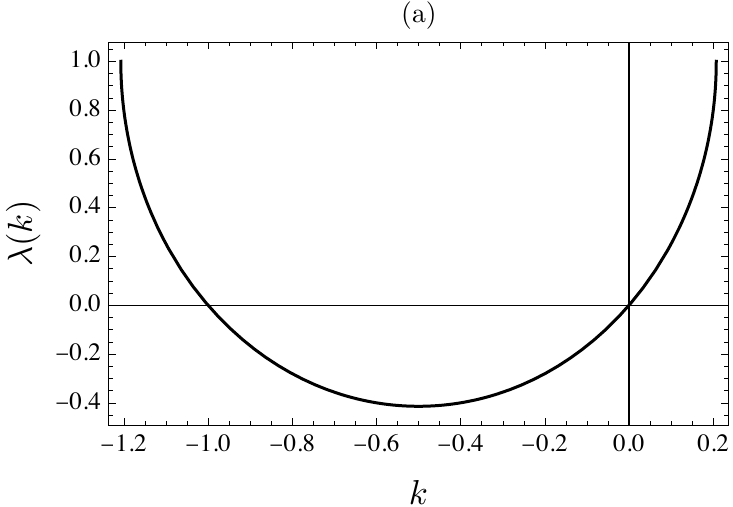}%
\hspace*{0.5in}
\includegraphics{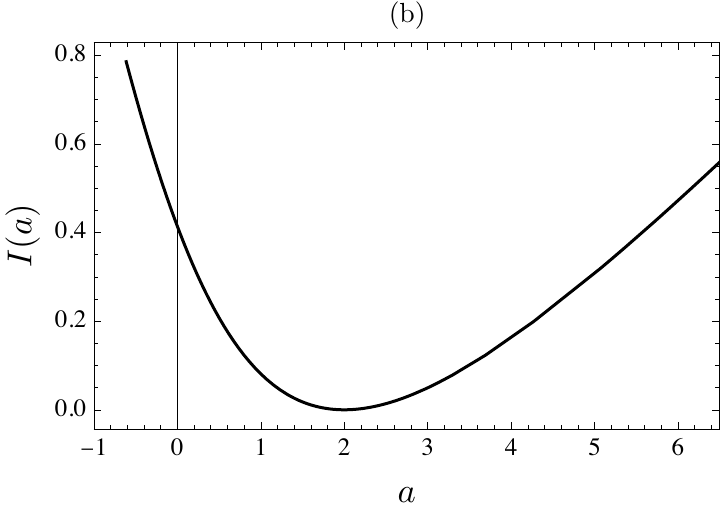}
\caption{(a) SCGF and (b) rate function of the nonequilibrium work done by the transverse diffusion for the parameters $\gamma = 1,\xi=1$, and $\epsilon=1$.}
\label{fig:transwork1}
\end{figure*}

We plot this function in Fig.~\ref{fig:transwork1}, together with the corresponding rate function $I(w)$ obtained by Legendre transform. The latter function has a minimum located at $w^*= \lambda'(0) = 2\xi^2/\gamma$ and has two branches on either side that become asymptotically linear in $w$, because the SCGF is defined on a bounded interval, which implies that the probability density $p(\work=w)$ has exponential tails for large work values, positive or negative. From the expression of the SCGF, we also note that
\be
\lambda(k) = \lambda(-k-1),
\label{fr1}
\ee
which is an important symmetry of the SCGF, referred to as the Gallavotti--Cohen fluctuation relation \cite{gallavotti1995,kurchan1998,lebowitz1999,harris2007}, which translates at the level of the rate function to
\be
I(w) = I(-w)-w.
\label{fr2}
\ee
Therefore, we have
\be
\frac{p(\work = w)}{p(\work = -w)} \approx e^{Tw}
\label{FR3}
\ee
for large $T$, which is the more standard expression of the Gallavotti--Cohen fluctuation relation, showing that positive work values are exponentially more likely than negative work values. This reflects the fact that the average work $w^*$ is always positive, since trajectories of the transverse diffusion travel on average in the direction of the rotating drift and, therefore, that negative work fluctuations resulting from trajectories that go against the drift are exponentially unlikely.

To understand how these fluctuations are created, we calculate the solution \eqref{mkcur} for the modified drift matrix using  the solution of the Riccati equation leading to the SCGF, obtaining
\be
M_k = 
\begin{pmatrix}
\gamma -\lambda(k) && \xi(1 + 2k) \\ 
-\xi(1 + 2k) && \gamma-\lambda(k)
\end{pmatrix}.
\ee
For the range $\cK^-$ above, this matrix is positive definite and so describes an ergodic effective process whose stationary density is
\be
p_k^*(\bx) = \frac{\gamma-\lambda(k)}{\pi\epsilon^2} e^{-[\gamma-\lambda(k)]\| \bx\|^2/{\epsilon^2}},
\label{ptranswork}
\ee
while the stationary current is
\be
 \bJ^*_k(\bx) = \xi(1 + 2k) 
\begin{pmatrix} 
-x_2 \\ x_1 
\end{pmatrix}
p_k^*(\bx).
\ee

\begin{figure*}[t]
\centering
\includegraphics{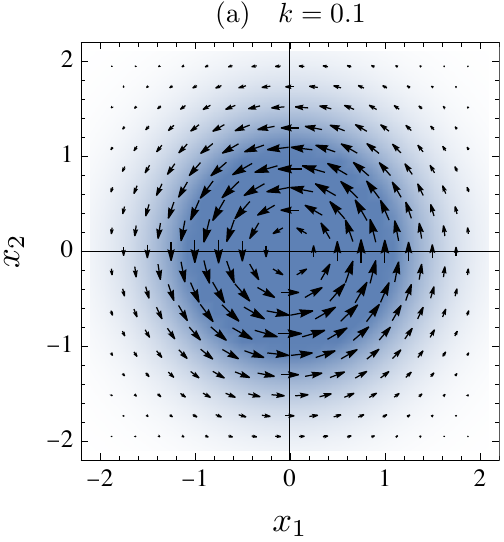}%
\hspace*{0.4cm}
\includegraphics{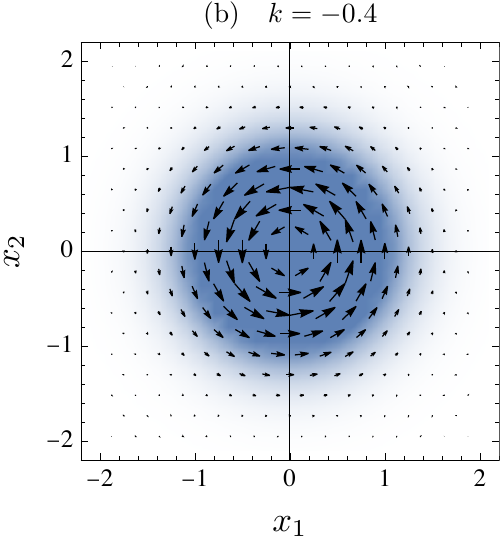}%
\hspace*{0.4cm}
\includegraphics{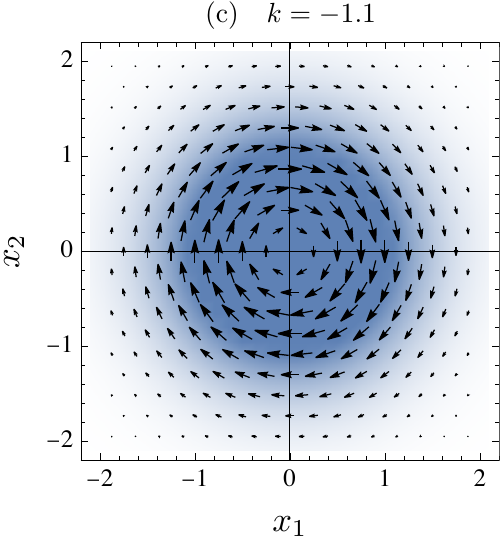}%
\caption{Vector plot of the stationary current $\bJ^*_k$ of the effective process associated with the nonequilibrium work done by the transverse system for different values of $k$. The density plots underneath show the modified stationary density $p^*_k$. Parameters: $\xi=1$, $\gamma =1$, and $\epsilon =1$.}
\label{fig:Jkwork1}
\end{figure*}

These are similar to the stationary density $p^*$ and current $\bJ^*$ found before in \eqref{rhotrans} and \eqref{Jtrans}, respectively, and are plotted in Fig.~\ref{fig:Jkwork1} for various values of $k$ and parameters values $\gamma=\epsilon=\xi=1$, giving rise to an anticlockwise $\bJ^*$. From the plots, we can see that positive work fluctuations are created by trajectories that have an anticlockwise current $\bJ^*_k$, as expected, which is greater in magnitude than $\bJ^*$ when $w>w^*$, corresponding to $k>0$ (see Fig.~\ref{fig:Jkwork1}a), and smaller in magnitude when $0<w<w^*$, corresponding to $-1/2 < k < 0$ (Fig.~\ref{fig:Jkwork1}b). On the other hand, for negative work fluctuations, associated with $k < -1/2$, the trajectories reverse direction (Fig.~\ref{fig:Jkwork1}c), thereby creating a clockwise current $\bJ^*_k$, which increases in magnitude as $k$ decreases. Between these two regimes, when $w=0$ (corresponding to $k = -1/2$), the current $\bJ^*_k$ vanishes, as the trajectories responsible for this work fluctuation do not rotate on average and behave, therefore, in a reversible way.

These changes in the current are also accompanied by changes in the density, as seen from \eqref{ptranswork}, which have the effect of confining the state either closer to the origin for $k\in (-1,0)$ (see Fig.~\ref{fig:Jkwork1}b) or further from it otherwise. Moreover, we can see that, as $k$ approaches the boundaries of $\cK^-$, the confinement, determined by diagonal part of $M_k$, vanishes, showing that the extremely large work fluctuations, either positive or negative, are effectively created by a weakly confined, rotating Brownian motion in the plane.

From these results, we can understand directly the large deviations of the entropy production by noting again that $\work$ is the antisymmetric part of $\ep$, so $\ep$ and $\work$ differ only by a boundary term, as discussed in the previous section. The boundary term, coming from the symmetric part of $\ep$, is described by the matrix
\be
\obsm^+ = -\frac{2\gamma}{\epsilon^2}\id,
\ee
which we use to determine the cut-off value beyond which the matrix $\cB_k$, defined in \eqref{posdef}, ceases to be positive definite. In our case, the cut-off is negative because $\obsm^+$ is negative definite and is equal to $k_{\min}=-1$, which means that the SCGF of $\ep$ matches that obtained for $\work$ but only for $k$ in the range
\be
\bigg[-1, \frac{-\xi^2 + \sqrt{\gamma^2 \xi^2+ \xi^4}}{2\xi^2}\bigg).
\ee

The effect of $k_{\min}$ on the rate function is similar to what we discussed in the previous section for $k_{\max}$ and leads here to the following rate function for $\ep$:
\be
I (e) = 
\begin{cases}
-e & e < \ce \\ 
I^-(e) & e \geq \ce,
\end{cases}
\ee
$I^-(e)$ being the rate function of the nonequilibrium work evaluated at arguments $\ep=e$ of the entropy production. The crossover value $\ce$ is determined from $\ce = \lambda'(-1)$ and is given explicitly by $\ce = -2\xi^2/\gamma$.

\begin{figure}[t]
\centering
\includegraphics{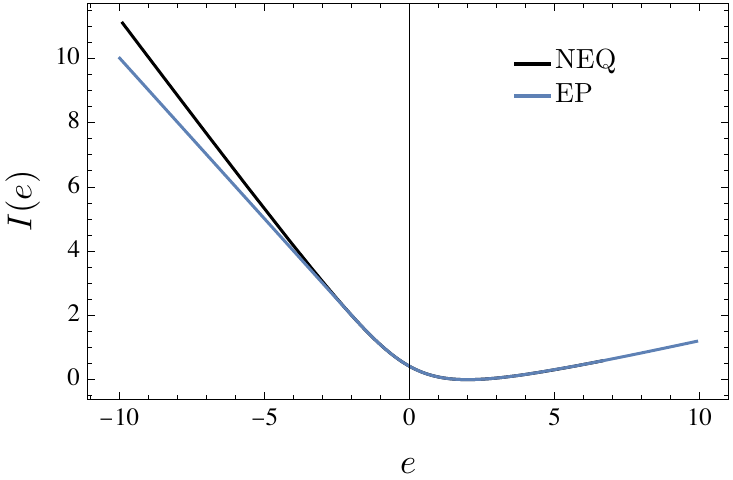}
\caption{Comparison of the rate functions of the nonequilibrium work (NEQ) and the entropy production (EP) for the transverse diffusion. Parameters: $\gamma = 1$, $\xi = 1$, and $\epsilon = 1$.}
\label{fig:ratecomp1}
\end{figure}

This rate function is compared with the rate function of the nonequilibrium work in Fig.~\ref{fig:ratecomp1}. The difference coming from the linear branch of $I(e)$ below $\ce$ is clearly seen and implies the existence of a dynamical phase transition that separates two large deviations regimes: one on the right of $\ce$ where the large deviations of $\ep$ are determined by the large deviations of $\work$, with the boundary term playing no role, and the other, left of $\ce$, where the large deviations of $\ep$ are only determined by those of the boundary term. The latter regime or region cannot be described in terms of an effective process, since it is related to the cut-off value $k_{\min}$. For $e>\ce$, however, there exists an effective process, which is the same for $\ep$ as for $\work$.

\subsection{Brownian gyrator}

We consider as our second application two Brownian particles with positions $X_1(t)$ and $X_2(t)$ evolving according to the overdamped SDE
\be
d\bX (t) = - 
\begin{pmatrix}
\gamma + \kappa && -\kappa \\ 
-\kappa && \gamma + \kappa 
\end{pmatrix}
\bX (t) + 
\begin{pmatrix}
\epsilon_1 && 0 \\ 
0 && \epsilon_2 
\end{pmatrix} 
d\bW (t),
\label{sdespring}
\ee
where $\bX (t) = (X_1(t),X_2(t))$ and $\bW(t) = (W_1(t), W_2(t))$. The drift in this system includes a friction force with friction parameter $\gamma>0$ and a linear (spring) force between the two particles with spring constant $\kappa\geq 0$. The presence of two separate noise strengths $\epsilon_1$ and $\epsilon_2$ indicates that the two particles interact with two different heat baths having, in general, non-identical temperatures $T_{1,2}= \epsilon_{1,2}^2/2$. The same SDE is also used to describe the charge dynamics of two resistors kept at different temperatures and coupled by a capacitance.

This system has been studied extensively in physics as the Brownian gyrator \cite{filliger2007,dotsenko2013,chiang2017}, and has a nonequilibrium steady state when $\kappa>0$ and $\epsilon_1\neq \epsilon_2$, related to the energy exchanged between the two thermal baths via the linear coupling. The stationary density and current characterizing this state can be calculated exactly, but their expressions are however too long to display here. For our purpose, we only note the stationary covariance matrix obtained from (\ref{ricccov}):
\begin{widetext}
\be
C = \frac{1}{4\gamma(\gamma + 2\kappa)} 
\begin{pmatrix}
\frac{2\gamma \epsilon_1^2 (\gamma + 2\kappa) + (\epsilon_1^2 + \epsilon_2^2)\kappa^2}{\gamma + \kappa} && (\epsilon_1^2 + \epsilon_2^2) \kappa \\ 
(\epsilon_1^2 + \epsilon_2^2) \kappa && \frac{2\gamma \epsilon_2^2 (\gamma + 2\kappa) + (\epsilon_1^2 + \epsilon_2^2)\kappa^2}{\gamma + \kappa} 
\end{pmatrix},
\ee
from which the stationary density and current can easily be found via (\ref{statden}) and (\ref{statcurlin}), respectively. It can be checked from this result that, if $\kappa > 0$ and the noise strengths are different, then a non-zero stationary probability current exists, which rotates clockwise in the plane when $\epsilon_1<\epsilon_2$ and anticlockwise when $\epsilon_1>\epsilon_2$. On the other hand, if $\epsilon_1=\epsilon_2$, then the system has an equilibrium steady state for arbitrary $\kappa$. Likewise, for $\kappa = 0$ the system is in equilibrium even when the noise strengths are different because the two particles are then decoupled, representing two isolated systems in contact with separate heat baths. 

For this system, we consider as before the nonequilibrium work $\work$, defined in \eqref{neqwork}, which was studied implicitly by Kwon \emph{et al.} \cite{kwon2011} and more recently by Monthus and Mazzolo \cite{monthus2022} using path integrals. This observable is characterized by the antisymmetric matrix
\be
\obsm^- = 
\begin{pmatrix}
0 && -\frac{\kappa(\epsilon_1^2 - \epsilon_2^2)}{\epsilon_1^2 \epsilon_2^2} \\ 
\frac{\kappa(\epsilon_1^2 - \epsilon_2^2)}{\epsilon_1^2 \epsilon_2^2} && 0
\end{pmatrix}. 
\ee
The SCGF cannot be found now by obtaining the generating function exactly, since $\ricmat_k(t)$ in the Riccati equation (\ref{riccatidiffcur}) does not have a diagonal form here, due to the off-diagonal symmetric part of the drift matrix $M$. However, we can solve the algebraic Riccati equation (\ref{ricccur}) so as to find the appropriate stationary solution $\ricmat^*_k$, leading to 
\be
\lambda(k) = \gamma + \kappa - \sqrt{\gamma^2 + 2\gamma \kappa - \kappa^2 \frac{\left((1+k)\epsilon_1^2 - k\epsilon^2 \right) \left(k \epsilon_1^2 - (1+k)\epsilon^2\right)}{\epsilon_1^2 \epsilon_2^2} }
\label{scgfspring}
\ee
for $k$ in the range $\cK^- = (k_-, k_+)$, where
\begin{equation}
k_{\pm} = 
\frac{-\kappa  \text{$\epsilon_1$}^2+\kappa  \text{$\epsilon_2$}^2\pm\sqrt{4 \gamma ^2 \text{$\epsilon_1 $}^2 \text{$\epsilon_2$}^2+8 \gamma  \kappa  \text{$\epsilon_1$}^2 \text{$\epsilon_2$}^2+\kappa ^2 \left(\text{$\epsilon_1$}^2+\text{$\epsilon_2$}^2\right)^2}}{2 \kappa  \left(\text{$\epsilon_1$}^2-\text{$\epsilon_2$}^2\right)}.
\label{krange4}
\end{equation}
\end{widetext}

\begin{figure*}[t]
\centering
\includegraphics{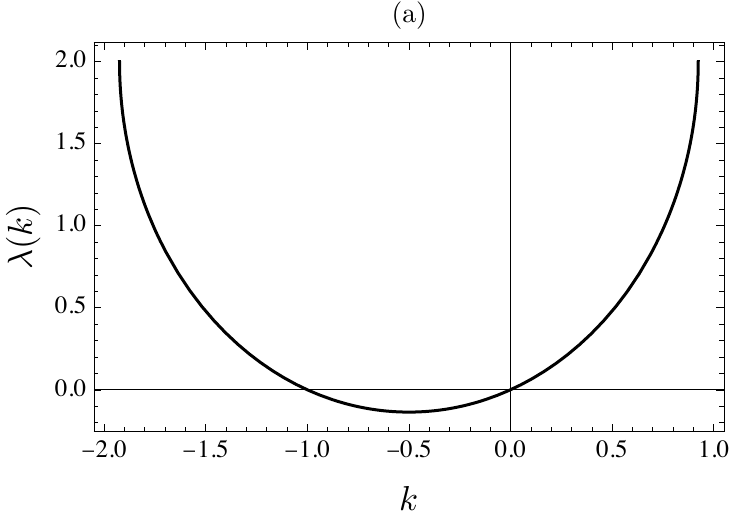}%
\hspace*{0.5in}
\includegraphics{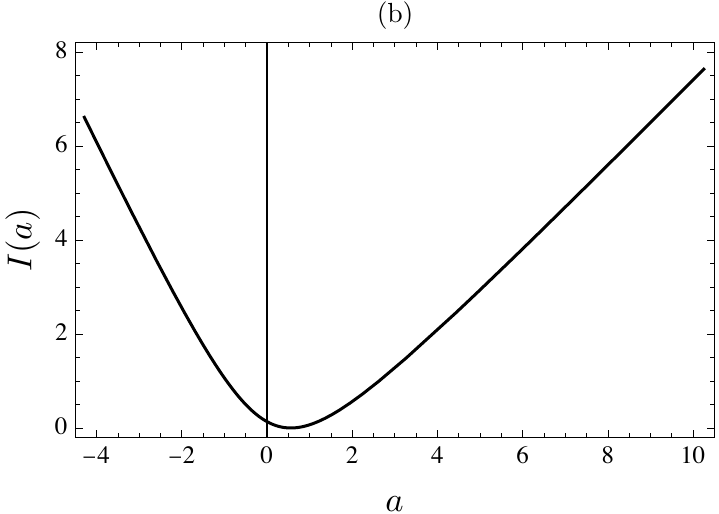}
\caption{(a) SCGF and (b) rate function of the nonequilibrium work done by Brownian gyrator for $\gamma = 1,\kappa=1$, and noise strengths $\epsilon_1 = 2$ and $\epsilon_2 =1$.}
\label{fig:springwork1}
\end{figure*}

This result is shown in Fig.~\ref{fig:springwork1} with the associated rate function, obtained by computing the Legendre transform numerically. The SCGF is symmetric around $k=-1/2$ and satisfies again the fluctuation symmetry noted before in \eqref{fr1}, which means that $I(w)$ satisfies the symmetry in \eqref{fr2}. The minimum of $I(w)$ is now located at 
\be
w^* = \frac{\kappa^2(\epsilon_1^2 - \epsilon_2^2)^2}{2(\gamma + \kappa)\epsilon_1^2 \epsilon_2^2},
\ee
predicting overall that positive work fluctuations are more likely than negative fluctuations with the same magnitude, in agreement with (\ref{FR3}). Further, it can be checked that $\lambda(k)$ and $I(w)$ remain invariant under the exchange $\epsilon_1 \leftrightarrow \epsilon_2$, indicating that only the magnitude of the difference $|\epsilon_1 - \epsilon_2|$ in noise strengths and not the sign of the difference $\epsilon_1 - \epsilon_2$ determines the large deviations. This is explained by noting that positive and negative values of $\work$ are determined by the direction or chirality of $\bJ^*$, as for the transverse diffusion, which depends here on the sign of $\epsilon_1 - \epsilon_2$.

\begin{figure*}[t]
\centering
\includegraphics{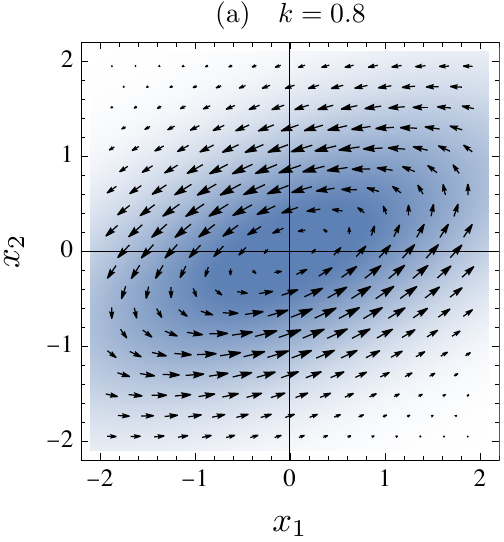}%
\hspace*{0.4cm}
\includegraphics{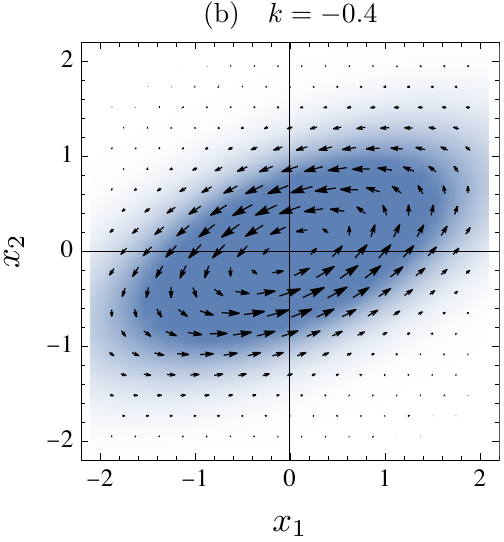}%
\hspace*{0.4cm}
\includegraphics{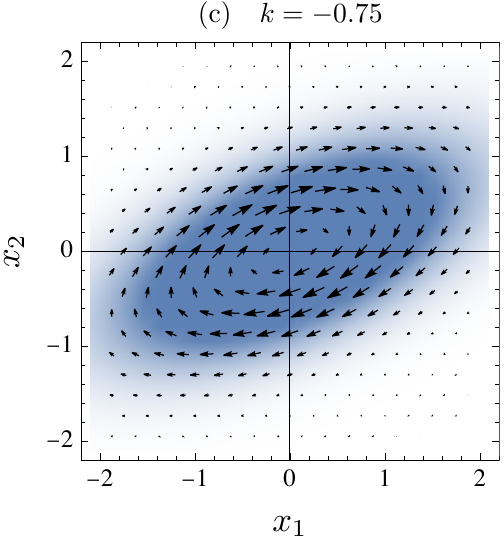}%
\caption{Vector plot of the stationary current $\bJ^*_k$ of the effective process associated with the nonequilibrium work done by the Brownian gyrator for various values of $k$. The density plots underneath show the modified stationary density $p^*_k$. Parameters: $\gamma =1$, $\kappa = 1$, $\epsilon_1 = 2$, and $\epsilon_2 = 1$.}
\label{fig:Jkspringwork}
\end{figure*}

The effective process underlying the fluctuations of $\work$ is similar to the one found for transverse diffusions, with $\epsilon_1 - \epsilon_2$ playing the role of the nonequilibrium parameter $\xi$, and so we do not discuss it in detail here. The main difference to note is that the stationary density $p^*$ of the Brownian gyrator has a tilt and eccentricity in the plane, related to the coupling $\kappa$, which are also seen in the stationary current. This property persists at the level of $p_k$ and $\bJ^*_k$ \footnote{It can be checked, in particular, that the matrix $H_k$ entering in the expression \eqref{Jquad} of $\bJ^*_k$ is proportional to $H$ for the Brownian gyrator, as is also the case for the transverse system \cite{buisson2022}. This seems to be a general property of linear currents.}, as shown in Fig.~\ref{fig:Jkspringwork}, but does not change otherwise the basic observation that positive work fluctuations follow the flow of the stationary current and affect only its magnitude (see Fig.~\ref{fig:Jkspringwork}a, b), while negative work fluctuations reverse the direction of the stationary current and also change its magnitude (Fig.~\ref{fig:Jkspringwork}c). For $\work=0$, which corresponds to $k=-1/2$, we also find $\bJ^*_k=\bs{0}$. In this case, the Brownian particles effectively cease to interact as they realize this work fluctuation, and thus behave in a reversible way.

\section{Conclusions}

We have studied in this paper the large deviations of linear SDEs, considering three types of dynamical observables, defined in terms of linear or quadratic integrals in time of the state. For these, we have obtained explicit formulas for the SCGF and rate function characterizing their probability distribution in the long-time limit. These formulas involve Riccati equations, which can be solved exactly in some cases, as illustrated here with two physically-motivated models, or numerically using methods developed in control theory \cite{bini2011}. In addition, we have studied how the fluctuations of these observables arise via rare trajectories that can be described in terms of an effective SDE, which includes extra terms in the drift driving the process in the fluctuation region of interest, or, equivalently, in terms of density and current fluctuations that differ from the stationary density and current of the SDE considered. These two complementary levels of fluctuations give valuable insights into how large deviations are created physically and show, for the three types of observables considered, that those large deviations originate from an effective SDE that is also linear. Consequently, they can be seen as arising from Gaussian density fluctuations coupled to current fluctuations that are both driven by linear non-conservative forces. 

In future studies, it would be interesting to study nonlinear SDEs and possibly nonlinear observables of these processes to see if useful information, exact or approximate, about their large deviations can be obtained by linearizing them in some way. For this problem, we see three applications of potential interest:
\begin{itemize}
\item Linearize the SDE and, if applicable, the observable near the fixed point of the noiseless dynamics, if there is one. Applying our results to the resulting linear model should describe the small Gaussian fluctuations of the actual nonlinear system and observable, meaning that the asymptotic mean and variance should be given by the linear model. 

\item The effective SDE associated with a nonlinear SDE and observable is, in general, another nonlinear SDE. In the case of quadratic observables and linear current-type observables, we expect both SDEs to have the same noiseless fixed point, if the original SDE has one, following what we have found for linear SDEs. Consequently, for these observables, we expect the linearized model to provide approximate information about the full range of large deviations. 

\item Many numerical and simulation methods rely on the knowledge of the effective process or attempt to construct that process in an iterative way in order to compute the SCGF or  the rate function \cite{chetrite2015,ferre2018,coghi2022,yan2022}. A linear ansatz could be included in these methods, either as an approximation of the effective process or as a seed for an iteration scheme that gradually constructs the correct nonlinear effective process. Both approaches could lead potentially to improved algorithms, since the spectral problem underlying the effective process would be replaced, effectively, by the problem of solving a Riccati equation.
\end{itemize} 

Other directions of interest include the generalization of our results to time-dependent linear diffusions, in particular, periodic linear diffusions, and to linear diffusions evolving in bounded domains with reflections at the boundaries. A framework for the large deviations of time-periodic systems has been developed \cite{barato2018} and application of this framework following the exact results obtained here for the generating function could prove fruitful. As for reflected diffusions, we have shown recently \cite{buisson2020b} that imposing reflecting boundaries to the simple one-dimensional Ornstein--Uhlenbeck process leads in general to a nonlinear effective process, because of additional boundary conditions imposed on the spectral problem \cite{mallmin2021}. It is therefore natural to ask how our results for unbounded linear diffusions, based on Riccati equations, are modified by these boundary conditions.

\begin{acknowledgments}
We thank Francesco Coghi and Raphael Chetrite for useful discussions. JdB is funded by the National Research Foundation, South Africa (PhD Scholarship).
\end{acknowledgments}

\bibliography{masterbibmin}

\end{document}